\documentclass{article}

\PassOptionsToPackage{square, numbers, compress}{natbib}

\usepackage[final]{NeurIPS}

\usepackage[utf8]{inputenc} 
\usepackage[T1]{fontenc} 
\usepackage{hyperref} 
\usepackage{url} 
\usepackage{booktabs} 
\usepackage{amsfonts} 
\usepackage{nicefrac} 
\usepackage{microtype} 
\usepackage{xcolor} 
\usepackage{multirow} 
\usepackage{graphicx, float, caption, subcaption} 

\definecolor{turquoise}{RGB}{0, 120, 200}

\hypersetup{
colorlinks = true,
linkcolor = turquoise,
citecolor = turquoise,
filecolor = turquoise,
urlcolor = turquoise,
bookmarks = true,
pdftoolbar = true,
pdfmenubar = true,
pdfstartview = {FitH},
pdftitle = {Automating Abnormality Detection in Musculoskeletal Radiographs through Deep Learning},
pdfauthor = {Goodarz Mehr},
pdfsubject = {MURA Deep Learning Model},
}

\title{Automating Abnormality Detection in Musculoskeletal Radiographs through Deep Learning}

\author{
	Goodarz Mehr\thanks{Graduate Research Assistant, Autonomous Systems and Intelligent Machines (ASIM) Lab} \\
	Department of Mechanical Engineering\\
	Virginia Polytechnic Institute and State University (Virginia Tech)\\
	Blacksburg, VA 24061 \\
	\texttt{goodarzm@vt.edu} \\
}

\begin{document}

	\maketitle
	
	\begin{abstract}
	
		This paper introduces MuRAD (Musculoskeletal Radiograph Abnormality Detection tool), a tool that can help radiologists automate the detection of abnormalities in musculoskeletal radiographs (bone X-rays). MuRAD utilizes a Convolutional Neural Network (CNN) that can accurately predict whether a bone X-ray is abnormal, and leverages Class Activation Map (CAM) to localize the abnormality in the image. MuRAD achieves an F1 score of 0.822 and a Cohen's kappa of 0.699, which is comparable to the performance of expert radiologists.
	
	\end{abstract}
	
	\section{Introduction} \label{Introduction}
	
	A critical task for radiologists is determining whether a study is abnormal. If the study is normal, the patient is free of disease and does not have to undergo further diagnostic procedures, but if it is not, the patient may need to be further evaluated if an abnormality is detected with some confidence. Musculoskeletal abnormality detection is particularly important as musculoskeletal conditions are the most common cause of disability and severe, long-term pain, with more than 1.7 billion people affected worldwide with over 30 million annual ER visits \cite{Woolf}. \par
	
	An automated tool for abnormality detection in bone X-rays has important clinical applications. First, it can be utilized for worklist prioritization. In this scenario, cases that are deemed abnormal can be prioritized so that the sickest patient gets the care first. Moreover, it can assign an initial reading of normal for studies without an abnormality, which can help manage workflow and, by providing rapid results to the ordering health provider, improve disposition in other areas of the healthcare system. Second, this tool can help combat radiologist fatigue. Radiologists worldwide are reading an increasing number of cases with more images per case, and problems are exacerbated for those in underserved communities \cite{Nakajima}. Fatigue can impact diagnostic accuracy, as evidenced by the results of a study that found there was a statistically significant reduction in fracture detection at the end of the work day compared to the beginning of the work day \cite{Krupinski}. A tool that can highlight the abnormality in an image can draw the attention of the radiologist, which could potentially reduce errors, speed up image interpretation, and help standardize diagnostic quality. \par
	
	A machine learning approach to this problem faces two challenges. The first challenge is data. In order to create a model that can, with high confidence, detect and localize abnormality in a bone X-ray, the model needs to be trained on a large collection of labeled musculoskeletal radiographs, which is hard to come by. The second challenge is finding the appropriate machine learning approach that can interpret the data with high accuracy. \par
	
	This paper introduces MuRAD (Musculoskeletal Radiograph Abnormality Detection tool) to overcome these challenges, a Convolutional Neural Network (CNN) trained on Stanford's MURA dataset and equipped with Class Activation Map (CAM) to determine whether an image is abnormal and localize the abnormality when it is detected \cite{Rajpurkar3}. Multiple CNNs were tested to find the one with the best performance, given that they allow us to adapt the model to the specific features of bone X-rays,. Furthermore, leveraging CAM helps us create heat maps on each image that highlight the detected abnormality. \par
	
	The remainder of this work is structured as follows. \autoref{Literature} provides a brief summary of the related work, \autoref{Method} describes our learning approach and the dataset used, \autoref{Results} discusses the results, and \autoref{Conclusion} concludes the findings of this paper.
	
	\section{Related work} \label{Literature}
	
	Large medical datasets have enabled expert-level performance for detection of hip fracture \cite{Gale}, pneumonia \cite{Rajpurkar1}, brain hemorrhage \cite{Grewal}, heart arrhythmia \cite{Rajpurkar2}, lymph node metastases \cite{Bejnordi}, skin cancer \cite{Esteva}, and diabetic retinopathy \cite{Gulshan}. There has been efforts to make medical radiography repositories available online, with a summary of the ones publicly available shown in \autoref{Datasets}. Except for MURA \cite{Rajpurkar4} (which is used in this study), all datasets have labels that are automatically generated from radiologist reports. \par
	
	\begin{table}[ht!]
	 	\caption{Overview of publicly available medical radiography image datasets \cite{Rajpurkar3}.}
		\label{Datasets}
		\centering
		\begin{tabular}{c c c c}
			\toprule
			dataset & Study type & Label & Images \\
			\midrule
			MURA \cite{Rajpurkar4} & musculoskeletal (upper extremity) & abnormality & 40,561 \\
			Pediatric Bone Age \cite{AIMI} & musculoskeletal (hand) & bone age & 14,236 \\
			Digital Hand Atlas \cite{Gertych} & musculoskeletal (left hand) & bone age & 1,390 \\
			ChestX-ray13 \cite{Wang} & chest & multiple pathologies & 112,120 \\
			OpenI \cite{Demner-Fushman} & chest & multiple pathologies & 7,470 \\
			MC \cite{Jaeger} & chest & abnormality & 138 \\
			Shenzhen \cite{Jaeger} & chest & Tuberculosis & 662 \\
			JSRT \cite{Shiraishi} & chest & pulmonary nodule & 247 \\
			DDSM \cite{Heath} & mammogram & breast cancer & 10,239 \\
			\bottomrule
		\end{tabular}
	\end{table}
	
	There are only a few publicly available datasets of musculoskeletal radiographs, and each of those contain fewer than 15,000 images. These include a dataset of pediatric hand radiographs annotated with skeletal age from Stanford Program for Artificial Intelligence in Medicine and Imaging \cite{AIMI}, and a dataset of left hand radiographs from children of different ages labeled with radiologist readings of bone age \cite{Gertych}. \par
	
	Along with the MURA dataset, which contains around 41,000 musculoskeletal radiographs from around 15,000 studies, researchers at Stanford developed a machine learning model for abnormality detection \cite{Rajpurkar3}. They trained a 169-layer CNN to estimate the probability of abnormality for each image in the study, with the network using a Dense Convolutional Network architecture \cite{Huang1}.
	
	\section{Method} \label{Method}
	
	For developing MuRAD, numerous CNNs were trained on the MURA dataset, and the one that performed best was equipped with CAM to locate abnormalities. The following subsections provide more detail about the dataset, model, and CAM implementation.
	
	\subsection{Dataset} \label{Section3.1}
	
	The Stanford MURA dataset was collected from HIPAA-compliant images from the Picture Archive and Communication System (PACS) of Stanford Hospital \cite{Rajpurkar3, Rajpurkar4}. The dataset consists of 40,561 multi-view radiographs from 14,863 studies of 12,173 patients between 2001 and 2012. Each image belongs to one of seven standard upper extremity radiographic study types: elbow, finger, forearm, hand, humerus, shoulder, and wrist. Each study was manually labeled as normal or abnormal by board-certified radiologists at the time of interpretation and diagnosis. The dataset is split into training (11,184 patients, 13,457 studies, 36,808 images), validation (783 patients, 1,199 studies, 3,197 images), and test (206 patients, 207 studies, 556 images) sets. A summary of the distribution of normal and abnormal studies in the training and validation sets is provided in \autoref{Data}.
	
	\begin{table}[ht!]
	 	\caption{Distribution of studies in the training and validation sets for the MURA dataset \cite{Rajpurkar3}.}
		\label{Data}
		\centering
		\begin{tabular}{c c c c c c}
			\toprule
			\multirow{2}{*}{Study} & \multicolumn{2}{c}{Train} & \multicolumn{2}{c}{Validation} & \multirow{2}{*}{Total} \\
			\cline{2 - 5}
			 & Normal & Abnormal & Normal & Abnormal & \\
			\midrule
			Elbow & 1,094 & 660 & 92 & 66 & 1,912 \\
			Finger & 1,280 & 655 & 92 & 83 & 2,110 \\
			Hand & 1,497 & 521 & 101 & 66 & 2,185 \\
			Humerus & 321 & 271 & 68 & 67 & 727 \\
			Forearm & 590 & 287 & 69 & 64 & 1,010 \\
			Shoulder & 1,364 & 1,457 & 99 & 95 & 3,015 \\
			Wrist & 2,134 & 1,326 & 140 & 97 & 3,697 \\
			\midrule
			Total number of studies & 8,280 & 5,177 & 661 & 538 & 14,656 \\
			\bottomrule
		\end{tabular}
	\end{table}
	
	The test set of the MURA dataset is not publicly available; therefore, in this study the results obtained from the validation set are reported instead.
	
	\subsection{Model} \label{Section3.2}
	
	Various CNNs were examined in order to select the one with the best performance for MuRAD. These models were variants of VGG \cite{Simonyan}, DenseNet \cite{Huang1}, ResNet \cite{He}, ResNeXt \cite{Xie}, Inception \cite{Szegedy}, and MobileNet \cite{Sandler}. A few of them are briefly described in what follows.
	
	The VGG model uses an architecture with very small (3 $\times$ 3) convolutional filters, which shows a significant improvement on previous configurations can be achieved by pushing the depth to 16 - 19 layers \cite{Simonyan}. The DenseNet model uses an architecture consisting of multiple dense blocks, in which each layer is connected to every other layer in a feed-forward fashion \cite{Huang1}. For each layer, the feature-maps of all preceding layers are used as inputs, and its own feature-maps are used as inputs into all subsequent layers. This architecture has several advantages: it alleviates the vanishing-gradient problem, strengthens feature propagation, encourages feature reuse, and substantially reduces the number of parameters. A DenseNet model can have depths as high as 201 layers. Finally, the ResNet model uses a residual learning framework for improving training of networks that are substantially deeper than those normally used \cite{He}. In this model each layer learns residual functions with reference to the layer inputs, instead of learning unreferenced functions. A ResNet model can have depths as high as 152 layers. \par
	
	Transfer learning concepts were leveraged for better training. The weights of each model were initialized to those of the model trained on the ImageNet dataset, which helped us reach the desired optimal point faster. Because of this, each image was normalized before training to have the same mean and standard deviation as those in the ImageNet dataset. Additionally, images were resized to 320 $\times$ 320 pixels and augmented by applying random lateral inversions and rotations of up to 30 degrees. Each model was trained for 100 epochs with mini-batches of size 16, 32, or 64 - depending on the model's use of GPU memory - using Adam with a learning rate of 0.0001 and default parameters. The learning rate was decayed by a factor of 10 each time the validation loss plateaued after an epoch. All models were trained on Virginia Tech's Advanced Research Computing NewRiver system. Each model was trained on one node utilizing two Nvidia P100 GPUs \cite{NewRiver}. \par
	
	After evaluating model performance, CAM was implemented on the best performing model. When inputting a radiograph $X$ into the fully trained model, if an abnormality was detected CAM would take a weighted average of the feature maps of the last convolutional layer to locate the abnormality. Essentially, if the $i$-th feature map output of the model on image $X$ was $f_{i}(X)$ and the $i$-th fully connected weight was $w_{i}$, CAM $M(X)$ could be written as
	\begin{equation} \label{CamEq}
		M(X) = \sum_{i} w_{i}f_{i}(X).
	\end{equation}
Finally, $M(X)$ was upscaled to the dimensions of the image and overlayed on the image to indicate the location of the abnormality.

	\section{Results} \label{Results}
	
	Overall, 18 CNNs were trained and their performance was evaluated on the validation dataset. These models were selected based on their ImageNet Top-1 crop error rate. A summary of the results can be seen in \autoref{OverallRes}.
	
	\renewcommand{\tabcolsep}{3.2 pt}
	\begin{table}[ht!]
	 	\caption{Performance comparison of CNN models trained on the MURA dataset.}
		\label{OverallRes}
		\centering
		\begin{tabular}{c c c c c c c c}
			\toprule
			\multirow{2}{*}{Model} & \multirow{2}{*}{Accuracy} & \multirow{2}{*}{Precision} & \multirow{2}{*}{Recall} & F1 & Cohen's & Number of & Training \\
			 &  &  &  & score & kappa & parameters (M) & time \\
			\midrule
			DenseNet 121 & 0.853 & 0.904 & \textbf{0.753} & 0.822 & 0.699 & 8 & 492 \\
			DenseNet 161 & 0.844 & 0.844 & 0.751 & 0.812 & 0.680 & 30 & 1083 \\
			DenseNet 169 & 0.843 & 0.896 & 0.736 & 0.808 & 0.678 & 14 & 595 \\
			DenseNet 201 & 0.847 & 0.899 & 0.742 & 0.813 & 0.685 & 20 & 888 \\
			Inception V3 & 0.787 & 0.854 & 0.632 & 0.727 & 0.558 & 25 & 354 \\
			MobileNet V2 & 0.837 & 0.896 & 0.721 & 0.799 & 0.665 & \textbf{3.4} & \textbf{203} \\
			ResNet 34 & 0.842 & 0.905 & 0.725 & 0.805 & 0.675 & 24.3 & 244 \\
			ResNet 50 & 0.838 & 0.910 & 0.710 & 0.798 & 0.666 & 25.6 & 427 \\
			ResNet 101 & 0.848 & 0.901 & 0.744 & 0.815 & 0.688 & 44.5 & 654 \\
			ResNet 152 & 0.852 & 0.913 & 0.742 & 0.819 & 0.696 & 60.2 & 1019 \\
			ResNext 50 & 0.844 & 0.893 & 0.742 & 0.810 & 0.680 & 25.6 & 1054 \\
			ResNext 101 & 0.842 & 0.896 & 0.734 & 0.807 & 0.676 & 44.5 & 2217 \\
			VGG 16 & 0.846 & \textbf{0.917} & 0.721 & 0.808 & 0.682 & 138 & 741 \\
			VGG 19 & 0.844 & 0.907 & 0.727 & 0.807 & 0.679 & 144 & 853 \\
			VGG 11 with BN$ ^ {*}$ & 0.837 & 0.896 & 0.719 & 0.798 & 0.663 & 133 & 491 \\
			VGG 13 with BN$ ^ {*}$ & 0.840 & 0.904 & 0.721 & 0.803 & 0.672 & 133 & 784 \\
			VGG 16 with BN$ ^ {*}$ & 0.850 & 0.911 & 0.738 & 0.815 & 0.691 & 138 & 954 \\
			VGG 19 with BN$ ^ {*}$ & \textbf{0.855} & 0.910 & 0.751 & \textbf{0.823} & \textbf{0.702} & 144 & 1087 \\
			\bottomrule
			$ ^ {*}$ batch normalization
		\end{tabular}
	\end{table}
	\renewcommand{\tabcolsep}{6 pt}
	
	\autoref{OverallRes} shows that DenseNet 121, ResNet 152, and VGG 19 with BN perform the best and Inception V3 performs the worst by a large margin. VGG 19 with BN scores the highest Cohen's kapp, along with the highest accuracy and F1 score. VGG 16 achieves the highest precision while DenseNet 121 scores the highest recall. Looking at training time, it is clear that MobileNet V2 and ResNet 34 are much faster to train compared to other models and both have a good performance. Overall, given that the three models with the best performance score virtually similar on all metrics but DenseNet 121 has far fewer parameters and faster to train, it is our model of choice. For mobile applications, MobileNet V2 can be a good replacement. \par
	
	The CNN trained by Stanford (DenseNet 169) scored an overall Cohen's kappa of 0.705 on the test set \cite{Rajpurkar3}. Furthermore, three expert radiologists scored an overall Cohen's kappa of 0.731, 0.763, 0.778 on the test set. While a direct comparison with those results is not possible, based on the performance of the trained DenseNet 169 model on the validation set we can make an educated guess that the three models with the highest performance will slightly surpass those results, perhaps scoring a Cohen's kappa of 0.72 on the test set. While the CNNs trained here have a better performance compared to those trained in similar studies \cite{Banga, Harini, Chakravarty, Shao1, Goyal, Shao2, Huang2}, they are still slightly behind expert human radiologists. \par
	
	\autoref{CohenRes} shows the Cohen's kappa values of six trained CNNs for different body parts. An overall look at the table shows that different models have different strengths and weaknesses depending on the category. DenseNet 121 performs the best for forearm and hand, MobileNet V2 for shoulder, ResNet 152 for finger and humerus, and VGG 19 with BN for elbow and wrist radiograph abnormality detection. These results suggest that an ensemble model that combines the best of these model to make category-based predictions could potentially have a better performance and even rival that of expert human radiologists. Finally, based on the results it seems that abnormality detection in hand, shoulder, and finger radiographs is more difficult than for the other body parts.
	
	\begin{table}[ht!]
	 	\caption{Cohen's kappa values of six trained models for different body parts.}
		\label{CohenRes}
		\centering
		\begin{tabular}{c c c c c c c}
			\toprule
			\multirow{2}{*}{Body part} & DenseNet & Inception & MobileNet & ResNet & ResNeXt & VGG 19 \\
			 & 121 & V3 & V2 & 152 & 50 & with BN \\
			\midrule
			Elbow & 0.735 & 0.718 & 0.734 & 0.734 & 0.748 & \textbf{0.761} \\
			Finger & 0.630 & 0.484 & 0.584 & \textbf{0.664} & 0.619 & 0.652 \\
			Forearm & \textbf{0.727} & 0.495 & 0.681 & 0.711 & 0.666 & 0.696 \\
			Hand & \textbf{0.587} & 0.363 & 0.540 & 0.543 & 0.570 & 0.570 \\
			Humerus & 0.807 & 0.615 & 0.748 & \textbf{0.822} & 0.793 & 0.807 \\
			Shoulder & 0.638 & 0.505 & \textbf{0.649} & 0.628 & 0.628 & 0.639 \\
			Wrist & 0.758 & 0.558 & 0.703 & 0.756 & 0.719 & \textbf{0.768} \\
			\bottomrule
		\end{tabular}
	\end{table}
	
\begin{figure}[ht!]	
		\centering
		\begin{subfigure}[h]{0.46\columnwidth}
			\centering
			\includegraphics[width = 0.84\columnwidth]{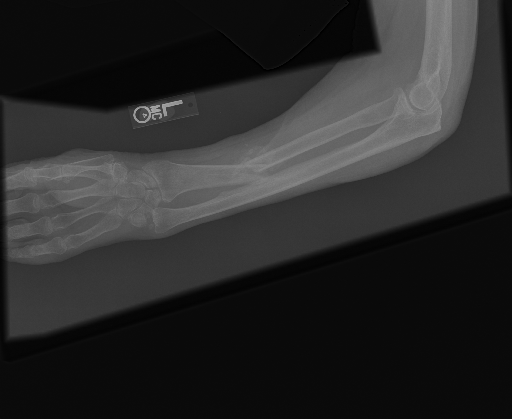}
			\includegraphics[width = 0.84\columnwidth]{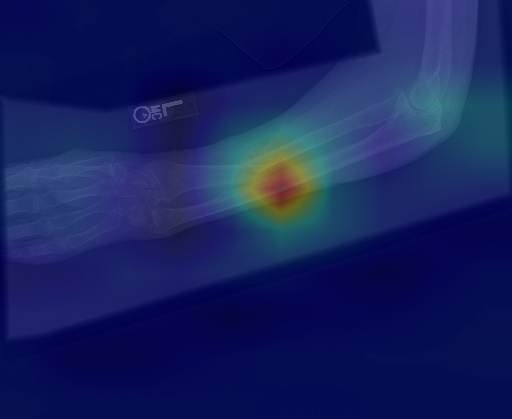}
			\caption{MuRAD detects a fracture in the forearm} \label{CamRadioA}
		\end{subfigure}
		\begin{subfigure}[h]{0.52\columnwidth}
			\centering
			\includegraphics[width = 0.608\columnwidth]{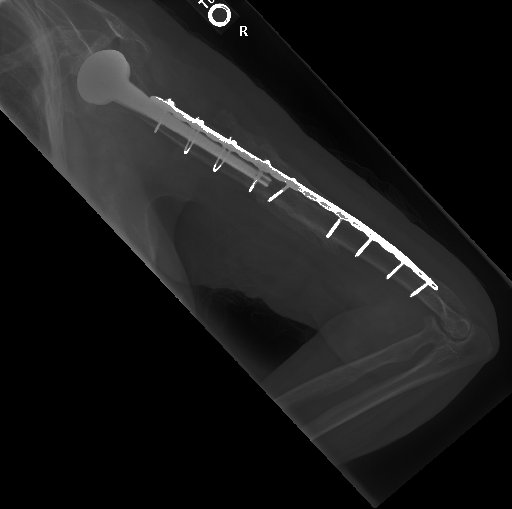}
			\includegraphics[width = 0.608\columnwidth]{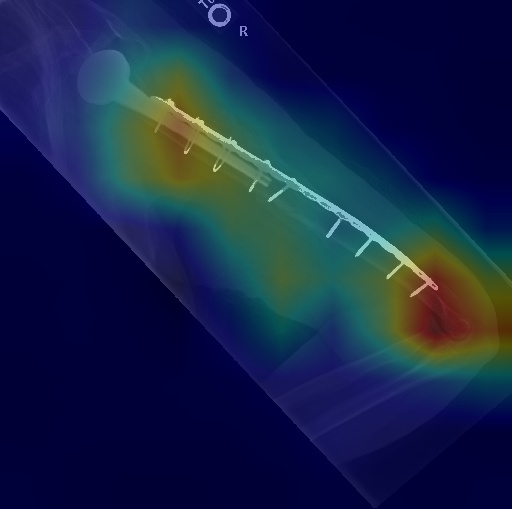}
			\caption{MuRAD detects a connecting rod and pins in the arm} \label{CamRadioB}
		\end{subfigure}
		\begin{subfigure}[h]{\columnwidth}
			\centering
			\includegraphics[width = 0.28\columnwidth]{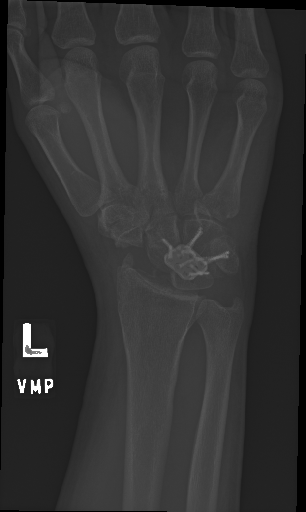}
			\includegraphics[width = 0.28\columnwidth]{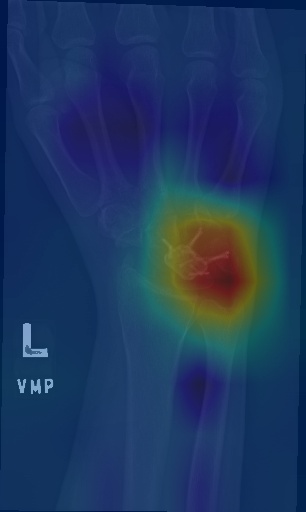}
			\caption{MuRAD detects a joint stabilizer in the wrist} \label{CamRadioC}
		\end{subfigure}
		\caption{Sample radiographs showcasing the performance of the CAM-equipped MuRAD.} \label{CamRadio}
	\end{figure}

	In the end, some sample radiographs showcasing the performance of MuRAD equipped with CAM are shown in \autoref{CamRadio}. As can be seen from the figure, CAM helps MuRAD accurately locate abnormalities in each image.
	
	\section{Conclusions} \label{Conclusion}
	
	This paper introduced MuRAD, a tool that can help radiologists automate the detection of abnormalities in musculoskeletal radiographs (bone X-rays). MuRAD utilizes a CNN that can accurately predict whether a bone X-ray is abnormal, and leverages CAM to localize the abnormality in the image. To develop MuRAD, 18 CNNs were trained on the Stanford MURA dataset and their performance was evaluated on the validation set. Out of these models, DenseNet 121 showed the best combination of performance, training time, and model size (number of parameters). Even though a direct comparison with the Stanford baseline model (DenseNet 169) was not possible due to the unavailability of the test set, the results indicate that MuRAD shows comparable performance to that model but is still slightly behind expert human radiologists. However, employing ensemble methods in a future work may narrow or completely eliminate this gap.
	
	\subsection*{Acknowledgments}
	
	The author would like to express his gratitude to the Stanford Machine Learning Group for making the MURA dataset publicly available, and to Virginia Tech's Advanced Research Computing center for providing the computational resources that made this work possible. This research was conducted using the PyTorch framework.

	\bibliographystyle{unsrt}
	\bibliography{MuraBib}

\end{document}